\input prepictex
\input pictex
\input postpictex
\typein[\sorb]{Enter ``s'' (for small) or ``b'' (for big)}
\newlength{\absize}

\if s\sorb \documentstyle{article}
\renewcommand{\baselinestretch}{1.5}
\renewcommand{\arraystretch}{1.5}
\begin{document}
\date{}
\pagestyle{empty}
\thispagestyle{empty}
\renewcommand{\thefootnote}{\fnsymbol{footnote}}
\newcommand{\starttext}{\newpage\normalsize
\pagestyle{plain}
\setlength{\baselineskip}{4ex}\par
\twocolumn\setcounter{footnote}{0}
\renewcommand{\thefootnote}{\arabic{footnote}}
}
\else
\documentstyle[12pt]{article}
\setlength{\absize}{6in}
\setlength{\topmargin}{-.5in}
\setlength{\oddsidemargin}{-.3in}
\setlength{\evensidemargin}{-.3in}
\setlength{\textheight}{9in}
\setlength{\textwidth}{7in}
\renewcommand{\baselinestretch}{1.5}
\renewcommand{\arraystretch}{1.5}
\setlength{\footnotesep}{\baselinestretch\baselineskip}
\begin{document}
\thispagestyle{empty}
\pagestyle{empty}
\renewcommand{\thefootnote}{\fnsymbol{footnote}}
\newcommand{\starttext}{\newpage\normalsize
\pagestyle{plain}
\setlength{\baselineskip}{4ex}\par
\setcounter{footnote}{0}
\renewcommand{\thefootnote}{\arabic{footnote}}
}
\fi

\newcommand{\preprint}[1]{\begin{flushright}
\setlength{\baselineskip}{3ex}#1\end{flushright}}
\renewcommand{\title}[1]{\begin{center}\LARGE
#1\end{center}\par}
\renewcommand{\author}[1]{\vspace{2ex}{\Large\begin{center}
\setlength{\baselineskip}{3ex}#1\par\end{center}}}
\renewcommand{\thanks}[1]{\footnote{#1}}
\renewcommand{\abstract}[1]{\vspace{2ex}\normalsize\begin{center}
\centerline{\bf Abstract}\par\vspace{2ex}\parbox{\absize}{#1
\setlength{\baselineskip}{2.5ex}\par}
\end{center}}

\newcommand{\segment}[2]{\put#1{\circle*{2}}}
\newcommand{\fig}[1]{figure~\ref{#1}}
\newcommand{\hc}{{\rm h.c.}}
\newcommand{\ds}{\displaystyle}
\newcommand{\eqr}[1]{(\ref{#1})}
\newcommand{\tr}{\,{\rm tr}}
\newcommand{\uone}{{U(1)}}
\newcommand{\su}[1]{{SU(#1)}}
\newcommand{\stu}{\su2\times\uone}
\newcommand{\be}{\begin{equation}}
\newcommand{\ee}{\end{equation}}
\newcommand{\bp}{\begin{picture}}
\newcommand{\ep}{\end{picture}}
\def\spur#1{\mathord{\not\mathrel{#1}}}
\def\lte{\mathrel{\displaystyle\mathop{\kern 0pt <}_{\raise .3ex
\hbox{$\sim$}}}}
\def\gte{\mathrel{\displaystyle\mathop{\kern 0pt >}_{\raise .3ex
\hbox{$\sim$}}}}
\newcommand{\sechead}[1]{\medskip{\bf #1}\par\bigskip}
\newcommand{\ba}[1]{\begin{array}{#1}\ds }
\newcommand{\cra}{\\ \ds}
\newcommand{\ea}{\end{array}}
\newcommand{\forto}[3]{\;{\rm for}\; #1 = #2 \;{\rm to}\; #3}
\newcommand{\for}{\;{\rm for}\;}
\newcommand{\cross }{\hbox{$\times$}}
\newcommand{\ol}{\overline}
\newcommand{\bra}[1]{\left\langle #1 \right|}
\newcommand{\ket}[1]{\left| #1 \right\rangle}
\newcommand{\braket}[2]{\left\langle #1 \left|#2\right\rangle\right.}
\newcommand{\braketr}[2]{\left.\left\langle #1 right|#2\right\rangle}
\newcommand{\g}[1]{\gamma_{#1}}
\newcommand{\half}{{1\over 2}}
\newcommand{\del}{\partial}
\newcommand{\grad}{\vec\del}
\newcommand{\real}{{\rm Re\,}}
\newcommand{\imag}{{\rm Im\,}}
\def\lta{\mathrel{\displaystyle\mathop{\kern 0pt <}_{\raise .3ex
\hbox{$\sim$}}}}
\def\gta{\mathrel{\displaystyle\mathop{\kern 0pt >}_{\raise .3ex
\hbox{$\sim$}}}}
\newcommand{\cl}[1]{\begin{center} #1\end{center}}
\newcommand\etal{{\it et al.}}
\newcommand{\prl}[3]{Phys. Rev. Letters {\bf #1} (#2) #3}
\newcommand{\prd}[3]{Phys. Rev. {\bf D#1} (#2) #3}
\newcommand{\npb}[3]{Nucl. Phys. {\bf B#1} (#2) #3}
\newcommand{\plb}[3]{Phys. Lett. {\bf #1B} (#2) #3}
\newcommand{\ie}{{\it i.e.}}
\newcommand{\etc}{{\it etc.\/}}
\def\cA{{\cal A}}
\def\cB{{\cal B}}
\def\cC{{\cal C}}
\def\cD{{\cal D}}
\def\cE{{\cal E}}
\def\cF{{\cal F}}
\def\cG{{\cal G}}
\def\cH{{\cal H}}
\def\cI{{\cal I}}
\def\cJ{{\cal J}}
\def\cK{{\cal K}}
\def\cL{{\cal L}}
\def\cM{{\cal M}}
\def\cN{{\cal N}}
\def\cO{{\cal O}}
\def\cP{{\cal P}}
\def\cQ{{\cal Q}}
\def\cR{{\cal R}}
\def\cS{{\cal S}}
\def\cT{{\cal T}}
\def\cU{{\cal U}}
\def\cV{{\cal V}}
\def\cW{{\cal W}}
\def\cX{{\cal X}}
\def\cY{{\cal Y}}
\def\cZ{{\cal Z}}
\renewcommand{\baselinestretch}{1.5}
\renewcommand{\arraystretch}{1.5}
\unitlength=1.2\unitlength
\newcommand{\BM}{\mbox{\boldmath$M$}}
\newsavebox{\phru}
\savebox{\phru}{\beginpicture
\setcoordinatesystem units <\unitlength,\unitlength>
\setquadratic
\plot
0 0
2.5 3
5 0
7.5 -3
10 0
/
\endpicture}
\newcommand{\photonru}[1]{\multiput(0,0)(10,0){#1}{\usebox{\phru}}}
\newsavebox{\phrd}
\savebox{\phrd}{\beginpicture
\setcoordinatesystem units <\unitlength,\unitlength>
\setquadratic
\plot
0 0
2.5 -3
5 0
7.5 3
10 0
/
\endpicture}
\newcommand{\photonrd}[1]{\multiput(0,0)(10,0){#1}{\usebox{\phrd}}}
\newsavebox{\phdr}
\savebox{\phdr}{\beginpicture
\setcoordinatesystem units <\unitlength,\unitlength>
\setquadratic
\plot
0 0
3 -2.5
0 -5
-3 -7.5
0 -10
/
\endpicture}
\newcommand{\photondr}[1]{\multiput(0,0)(0,-10){#1}{\usebox{\phdr}}}
\newsavebox{\phdl}
\savebox{\phdl}{\beginpicture
\setcoordinatesystem units <\unitlength,\unitlength>
\setquadratic
\plot
0 0
-3 -2.5
0 -5
3 -7.5
0 -10
/
\endpicture}
\newcommand{\photondl}[1]{\multiput(0,0)(0,-10){#1}{\usebox{\phdl}}}
\newcommand{\ptcommand}[1]{\put(0,0){\beginpicture
\setcoordinatesystem units <\unitlength,\unitlength>
#1
\endpicture}}
\setlength{\parindent}{3em}
\setlength{\footnotesep}{.6\baselineskip}
\newcommand{\myfoot}[1]{\footnote{\setlength{\baselineskip}{.75\baselineskip}
#1}}
\renewcommand{\thepage}{\arabic{page}}
\setcounter{bottomnumber}{2}
\setcounter{topnumber}{3}
\setcounter{totalnumber}{4}
\newcommand{\figsize}{}
\renewcommand{\bottomfraction}{1}
\renewcommand{\topfraction}{1}
\renewcommand{\textfraction}{0}
\newcommand{\ploops}{+{\mbox{\small loop}\atop\mbox{\small effects}}}
\preprint{\#HUTP-92/A037\\ 9/92}
\def\theequation{\thesection.\arabic{equation}}
\title{Physics from Vacuum Alignment in\\
A Technicolor Model\thanks{Research
supported in part by the National Science Foundation under Grant
\#PHY-8714654.}\thanks{Research supported in part by the Texas National
Research Laboratory Commission, under Grant \#RGFY9206.}
}
\author{
Howard Georgi \\
Lyman Laboratory of Physics \\
Harvard University \\
Cambridge, MA 02138
}
\date{}
\abstract{
I describe a technicolor model (without walking) of all interactions
up to energies of the order of 1000~TeV. The low energy extended technicolor
model has the CTSM form. But at intermediate energy scales, additional
interactions are required to give the KM mixing. I discuss the characteristic
flavor changing neutral current interactions arising from this flavor physics
below 1000~TeV.}

\starttext
\section{Introduction}

In CTSM models,~\cite{ctsm,ctsmquarks,nagoya} the flavor changing neutral
current (FCNC) problem of conventional extended technicolor (ETC) models is
solved by the introduction of three different ETC groups. This allows us to
keep the GIM~\cite{gim} symmetry structure of the standard model. The GIM
symmetries are broken by the masses of heavy, strongly interacting
``ultrafermions''. The masses come from 4-fermion operators which come from
unspecified physics at a large scale, large enough that FCNC effects from this
scale are small.

In recent work, Lisa Randall reexamined this subject from a slightly different
point of view, that she calls the ``Extended Technicolor Standard Model,'' to
emphasize that compositeness is not a necessary ingredient for models of this
kind. She has also suggested a solution to the flavor anomaly
problem in CTSM models.~\cite{lisa} The flavor anomaly problem arises because
whenever the ultracolor group in a CTSM model is nonabelian, there must be
more than one light state carrying each flavor quantum number. Randall's
proposed solution is to assume that the flavor symmetries are broken
spontaneously at an intermediate mass scale (between the $\stu$ breaking scale
and the QCD scale). The flavor anomalies of the light particles are then
saturated by pseudoGoldstone bosons produced at this scale. The resulting
phenomenology is very rich and dangerous. The model that I discuss in this
note, incorporates a relatively benign version of Randall's idea, very similar
to the second model described in \cite{lisa}. In this
version, most of the pseudoGoldstone bosons can be very heavy, well out of
reach until
SSC energies. This is sad, because it makes these objects much harder to see.
However, I will not discuss this physics in any detail. My primary purpose is
to discuss some other aspects of CTSM models. In particular, I build a version
of the model of \cite{nagoya} that describes leptons as well as quarks, and
has only a single doublet of technifermions at the $\stu$ breaking scale. An
interesting feature of this model is that the KM matrix is determined, in
part, by a vacuum alignment problem. The discussion of \cite{ctsmquarks}
suggests that this feature is generic to realistic CTSM models. I investigate
the structure of these models in more detail, paying particular attention to
the additional flavor changing neutral currents that arise because of this
vacuum alignment. I argue that these flavor changing neutral current effects
may provide another window into CTSM.

I begin, in section \ref{moosesec}, by reviewing ``moose'' notation. In
section \ref{modelsec}, I describe the model, introduce the various scales of
new physics, and in section \ref{ctsmsec}, discuss the CTSM vacuum alignment.
In section \ref{kmsec}, I discuss the vacuum alignment that fixes the KM
matrix. In section \ref{etcsec}, I discuss the masses of the ETC gauge bosons.
These give rise to the quark and lepton masses, and also the new FCNC effects.
In section \ref{tsec}, I discuss the limit in which the ultrafermion fermion
corresponding to the $t$ quark is sufficiently heavy to affect the CTSM vacuum
alignment, so that the $t$ gets a mass by a slightly different mechanism. In
section \ref{scalesec}, I estimate the scales in the model. Section
\ref{lastsec} contains conclusions.

\section{Moose notation\label{moosesec}}

A note about ``moose'' notation \cite{moose}: The moose diagrams are a simple
way of writing down the gauge and fermion content of certain kinds of field
theories. Each circle is an $\su{n}$ gauge group, with $n$ and the label for
the gauge group indicated inside the circle. Each line between two circles is
a left-handed fermion representation, transforming like the defining
representation under the group from which the arrow is coming, and the complex
conjugate
representation under the group to which the arrow is pointing. Thus the moose
\begin{equation}
\bp(70,30)(-10,-5)
\put(0,0){\circle{20}}
\put(50,0){\makebox(0,0){$m\atop M$}}
\put(50,0){\circle{20}}
\put(0,0){\makebox(0,0){$n\atop N$}}
\put(10,0){\line(1,0){30}}
\put(10,0){\vector(1,0){15}}
\ep
\label{moose}
\end{equation}
represents a left-handed fermion transforming as an $n$ under the $\su{n}$
group, $N$, and as an $\ol{m}$ under the $\su{m}$ group, $M$. A more
conventional representation of this would be $(n,\ol{m})_L$. When we need to
refer to the fermion in a formula, we will label the left-handed fermion
descriptively as
\begin{equation}
\mbox{$[NM]_L$, an $n$$\times$$m$ matrix.}
\label{left}
\end{equation}
Under an $N$ transformation associated with an $n$$\times$$n$ unitary matrix,
$\cN$, and an $M$ transformation with $m$$\times$$m$ unitary matrix,, $\cM$,
this transforms as
\begin{equation}
[NM]_L\rightarrow \cN\,[NM]_L\,\cM^\dagger\,.
\label{lefttransform}
\end{equation}
The charge conjugate, right-handed fermion will be
\begin{equation}
\mbox{$[MN]_R$, an $m$$\times$$n$ matrix,}
\label{right}
\end{equation}
transforming as
\begin{equation}
[MN]_R\rightarrow \cM\,[MN]_R\,\cN^\dagger\,.
\label{righttransform}
\end{equation}
It is also convenient to define
\begin{equation}
\ol{[MN]_L}\equiv [NM]_L^\dagger\gamma^0\,,\quad
\ol{[NM]_R}\equiv [MN]_R^\dagger\gamma^0\,,
\label{bars}
\end{equation}
which transform appropriately,
\begin{equation}
\ba{c}
\ol{[MN]_L}\rightarrow \cM\,\ol{[MN]_L}\,\cN^\dagger\,,\\
\ol{[NM]_R}\rightarrow \cN\,\ol{[NM]_R}\,\cM^\dagger\,.\ea
\label{barstransform}
\end{equation}

\section{The model\label{modelsec}}

\begin{figure}[htb]$$
\bp(300,210)(-150,-50)
\multiput(-100,0)(50,0){5}{\circle{20}}
\multiput(0,50)(0,50){2}{\circle{20}}
\multiput(-110,0)(50,0){3}{\vector(-1,0){15}}
\multiput(-140,0)(50,0){3}{\line(1,0){30}}
\multiput(110,0)(-50,0){3}{\vector(1,0){15}}
\multiput(140,0)(-50,0){3}{\line(-1,0){30}}
\multiput(0,40)(0,50){3}{\vector(0,-1){15}}
\multiput(0,10)(0,50){3}{\line(0,1){30}}
\put(-100,0){\makebox(0,0){$n+k \atop U$}}
\put(100,0){\makebox(0,0){$n+k \atop D$}}
\put(0,100){\makebox(0,0){$n+k \atop L$}}
\put(0,40){\line(-1,-1){20}}
\put(0,40){\vector(-1,-1){10}}
\put(-50,-40){\vector(0,1){15}}
\put(-50,-10){\line(0,-1){30}}
\put(50,-40){\vector(0,1){15}}
\put(50,-10){\line(0,-1){30}}
\put(-50,0){\makebox(0,0){$s\atop X_U$}}
\put(50,0){\makebox(0,0){$s\atop X_D$}}
\put(0,50){\makebox(0,0){$2s\atop X_L$}}
\put(50,-50){\makebox(0,0){$k_D$}}
\put(-25,15){\makebox(0,0){$k_L$}}
\put(-50,-50){\makebox(0,0){$k_U$}}
\put(0,150){\makebox(0,0){2}}
\put(0,0){\makebox(0,0){$n\atop L_{TC}$}}
\put(-40,100){\vector(1,0){15}}
\put(-10,100){\line(-1,0){30}}
\put(-90,100){\vector(1,0){15}}
\put(-60,100){\line(-1,0){30}}
\put(-100,50){\makebox(0,0){$s-1\atop Y_U$}}
\put(-50,100){\makebox(0,0){$s-1\atop Z_U$}}
\put(40,100){\vector(-1,0){15}}
\put(10,100){\line(1,0){30}}
\put(90,100){\vector(-1,0){15}}
\put(60,100){\line(1,0){30}}
\put(-100,100){\circle{20}}
\put(100,100){\circle{20}}
\put(100,100){\makebox(0,0){$n\atop D'$}}
\put(50,50){\makebox(0,0){$k\atop D''$}}
\put(-100,100){\makebox(0,0){$n\atop U'$}}
\put(-50,50){\makebox(0,0){$k\atop U''$}}
\put(50,100){\circle{20}}
\put(100,50){\circle{20}}
\put(-50,100){\circle{20}}
\put(-100,50){\circle{20}}
\put(100,50){\makebox(0,0){$s-1\atop Y_D$}}
\put(50,100){\makebox(0,0){$s-1\atop Z_D$}}
\put(-50,50){\circle{20}}
\put(-40,100){\vector(1,0){15}}
\put(-10,100){\line(-1,0){30}}
\put(-90,100){\vector(1,0){15}}
\put(-60,100){\line(-1,0){30}}
\put(50,50){\circle{20}}
\put(90,50){\vector(-1,0){15}}
\put(60,50){\line(1,0){30}}
\put(-90,50){\vector(1,0){15}}
\put(-60,50){\line(-1,0){30}}
\put(100,60){\vector(0,1){15}}
\put(100,90){\line(0,-1){30}}
\put(50,60){\vector(0,1){15}}
\put(50,90){\line(0,-1){30}}
\put(100,10){\vector(0,1){15}}
\put(100,40){\line(0,-1){30}}
\put(-100,60){\vector(0,1){15}}
\put(-100,90){\line(0,-1){30}}
\put(-50,60){\vector(0,1){15}}
\put(-50,90){\line(0,-1){30}}
\put(-100,10){\vector(0,1){15}}
\put(-100,40){\line(0,-1){30}}
\ep
$$
\caption{\label{highmoose}The moose above the scale $f_1$.}
\end{figure}

Now consider the moose shown in \fig{highmoose}. This is a modification of the
model of \cite{nagoya}, incorporating a version of Randall's idea for dealing
with the flavor anomalies.\myfoot{I have not made any effort to deal with the
right handed neutrinos in a clever way because I want to concentrate on the
quark mixing. See reference \cite{lisa} for some possibilities.} We assume
that some unspecified high energy physics leaves us with 4-fermion operators
at a large scale $f_0$. For example, proportional to
\begin{equation}\ba{c}
\tr \left\{\cM_U\left(\ol{[k_UX_U]_R}\gamma_\mu[X_UL_{TC}]_R\right)
\left(\ol{[L_{TC}X_L]_L}\gamma^\mu[X_Lk_L]_L\right)\right\}\,,\\
\tr \left\{\cM_D\left(\ol{[k_DX_D]_R}\gamma_\mu[X_DL_{TC}]_R\right)
\left(\ol{[L_{TC}X_L]_L}\gamma^\mu[X_Lk_L]_L\right)\right\}\,.\ea
\label{fourfermionmasses}
\end{equation}

Next, $\alpha_{L_{TC}}$ gets strong at scale
$f_1$.\myfoot{\label{custodial}Note that if
$\alpha_{X_L}\ll\alpha_{X_U},\alpha_{X_D}$ in the high energy theory, then the
couplings $\alpha_{X_U}$ and $\alpha_{X_D}$ in the low energy theory are about
equal. Maybe this is a way of getting custodial $\su2$
naturally.}$^,$\myfoot{\label{ctsmfoot}Note also that this symmetry breaking
produces 4-fermion operators that help with the CTSM alignment. If the
couplings $\alpha_{X_U}$ and $\alpha_{X_D}$ are large, or there is a little
walking, this could be very important.} This gives the moose shown in
\fig{lowmoose}.

\begin{figure}[htb]$$
\bp(300,160)(-150,-50)
\multiput(-100,0)(50,0){5}{\circle{20}}
\multiput(-110,0)(50,0){3}{\vector(-1,0){15}}
\multiput(-140,0)(50,0){3}{\line(1,0){30}}
\multiput(110,0)(-50,0){3}{\vector(1,0){15}}
\multiput(140,0)(-50,0){3}{\line(-1,0){30}}
\multiput(0,40)(0,50){1}{\vector(0,-1){15}}
\multiput(0,10)(0,50){1}{\line(0,1){30}}
\put(-100,0){\makebox(0,0){$n+k \atop U$}}
\put(100,0){\makebox(0,0){$n+k \atop D$}}
\put(0,0){\makebox(0,0){$n+k \atop L$}}
\put(-50,0){\makebox(0,0){$s\atop X_U$}}
\put(50,0){\makebox(0,0){$s\atop X_D$}}
\put(0,50){\makebox(0,0){2}}
\put(-80,-35){\makebox(0,0){$k_U$}}
\put(-20,-35){\makebox(0,0){$k_{LU}$}}
\put(80,-35){\makebox(0,0){$k_D$}}
\put(20,-35){\makebox(0,0){$k_{LD}$}}
\put(-50,-40){\makebox(0,0){$\cM_U$}}
\put(50,-40){\makebox(0,0){$\cM_D$}}
\put(-80,-20){\vector(1,1){10}}
\put(-60,0){\line(-1,-1){20}}
\put(-40,0){\vector(1,-1){10}}
\put(-20,-20){\line(-1,1){20}}
\put(80,-20){\vector(-1,1){10}}
\put(60,0){\line(1,-1){20}}
\put(40,0){\vector(-1,-1){10}}
\put(20,-20){\line(1,1){20}}
\put(-90,100){\vector(1,0){15}}
\put(-60,100){\line(-1,0){30}}
\put(-100,50){\makebox(0,0){$s-1\atop Y_U$}}
\put(-50,100){\makebox(0,0){$s-1\atop Z_U$}}
\put(90,100){\vector(-1,0){15}}
\put(60,100){\line(1,0){30}}
\put(-100,100){\circle{20}}
\put(100,100){\circle{20}}
\put(100,100){\makebox(0,0){$n\atop D'$}}
\put(50,50){\makebox(0,0){$k\atop D''$}}
\put(-100,100){\makebox(0,0){$n\atop U'$}}
\put(-50,50){\makebox(0,0){$k\atop U''$}}
\put(50,100){\circle{20}}
\put(100,50){\circle{20}}
\put(-50,100){\circle{20}}
\put(-100,50){\circle{20}}
\put(100,50){\makebox(0,0){$s-1\atop Y_D$}}
\put(50,100){\makebox(0,0){$s-1\atop Z_D$}}
\put(-50,50){\circle{20}}
\put(-90,100){\vector(1,0){15}}
\put(-60,100){\line(-1,0){30}}
\put(50,50){\circle{20}}
\put(90,50){\vector(-1,0){15}}
\put(60,50){\line(1,0){30}}
\put(-90,50){\vector(1,0){15}}
\put(-60,50){\line(-1,0){30}}
\put(100,60){\vector(0,1){15}}
\put(100,90){\line(0,-1){30}}
\put(50,60){\vector(0,1){15}}
\put(50,90){\line(0,-1){30}}
\put(100,10){\vector(0,1){15}}
\put(100,40){\line(0,-1){30}}
\put(-100,60){\vector(0,1){15}}
\put(-100,90){\line(0,-1){30}}
\put(-50,60){\vector(0,1){15}}
\put(-50,90){\line(0,-1){30}}
\put(-100,10){\vector(0,1){15}}
\put(-100,40){\line(0,-1){30}}
\put(-45.5,91){\vector(1,-2){20.5}}
\put(-45.5,91){\line(1,-2){41}}
\put(45.5,91){\vector(-1,-2){20.5}}
\put(45.5,91){\line(-1,-2){41}}
\ep
$$\caption{\label{lowmoose}The moose below the scale $f_1.$}
\end{figure}

The 4-fermion operators of \eqr{fourfermionmasses} become mass
terms,\myfoot{Note that $[X_Lk_L]_L$ splits into $[X_Uk_{LU}]_L$ and
$[X_Dk_{LD}]_L$.}
\begin{equation}\ba{c}
\tr \left\{\cM_U\left(\ol{[k_UX_U]_R}\,[X_Uk_{LU}]_L\right)\right\}\,,\\
\tr \left\{\cM_D\left(\ol{[k_DX_D]_R}\,[X_Dk_{LD}]_L\right)\right\}\,.\ea
\label{fermionmasses}
\end{equation}

In addition, this leaves 4-fermion operators coupling $k_{LU}$ to $k_{LD}$
with coefficient of order
\begin{equation} {1\over f_1^2}\,. \label{2}\end{equation}
Below we estimate the sizes of things using a slightly elaborated version of
naive dimensional analysis (NDA),~\cite{newnda} in which we introduce for each
group a Goldstone boson decay constant, $f$, and a mass scale, $\Lambda$. It
is the decay constant, $f_1$, that enters into \eqr{2}, because
these operators arise from exchange of heavy gauge bosons produced at the
scale $f_1$, as shown in \fig{moose14}, where the double wavy line represents
the heavy gauge boson. Thus \eqr{2} is just the perturbative result.

\begin{figure}
$$
\bp(80,80)(-40,-30)
\put(-40,20){\line(1,0){80}}
\put(-40,20){\vector(1,0){20}}
\put(0,20){\vector(1,0){20}}
\put(-40,-20){\line(1,0){80}}
\put(40,-20){\vector(-1,0){20}}
\put(0,-20){\vector(-1,0){20}}
\put(-2,20){\photondr{4}}
\put(2,20){\photondr{4}}
\put(-20,25){\makebox(0,0)[rb]{$[X_Uk_{LU}]_L$}}
\put(-20,-25){\makebox(0,0)[rt]{$[X_Uk_{LU}]_L$}}
\put(20,25){\makebox(0,0)[lb]{$[X_Dk_{LD}]_L$}}
\put(20,-25){\makebox(0,0)[lt]{$[X_Dk_{LD}]_L$}}
\ep
$$
\caption{\label{moose14}The 4-fermion operator coupling $k_{LU}$ and
$k_{LD}$.}
\end{figure}

Next $\alpha_{X_U}$ and $\alpha_{X_D}$ get strong at scales
$f_{2U}$ and $f_{2D}$, with typical CTSM pattern. As we will discuss in more
detail in section \ref{ctsmsec}, this is only possible for
\begin{equation} {\cM_U\over\Lambda_{2U}}\lta{\xi_U\xi_L}\,,\quad
{\cM_D\over\Lambda_{2D}}\lta{\xi_D\xi_L}
\,.\label{3}\end{equation}
We have defined
\begin{equation}\xi_j\equiv{\alpha_j\over4\pi}\,.\label{xi}\end{equation}

Assume (see footnote \ref{custodial})
\begin{equation} f_{2U}\approx f_{2D}\equiv f_2
\,,\quad \Lambda_{2U}\approx \Lambda_{2D}
\equiv \Lambda_2\,,\label{4}\end{equation}
and consider the vacuum alignment. We will do this completely ignoring the
fermions transforming under the $Y$ and $Z$ groups. This is the sector of the
theory discussed by Randall.~\cite{lisa} The only innovation here is split up
the $U'$ and $U''$ and $D'$ and $D''$ groups. This is important to avoid large
contributions from this physics to the light fermion mass, but need not
concern us here. If these groups trigger symmetry breaking at a scale somewhat
below $f_2$, they are not very important in the analysis of the structure of
the theory at $f_2$.

The $k_U$ and $k_D$ condensates can be put
in canonical form by $U$ and $D$ gauge transformations, as shown below in
\eqr{xiu} and \eqr{xid}. But the $k_{LU}$ and $k_{LD}$ condensates are
coupled. The product of the two, invariant under $L$, is a unitary matrix,
$U$.  Then the part of the potential that determines
$U$ looks like
\begin{equation} -{af_2\over \Lambda_2^2}{\xi_L^2}
\tr\left(\cM_U\cM_U^\dagger U
\cM_D\cM_D^\dagger U^\dagger\right)
-{\xi_L}{16\pi^2f_2^6b\over f_1^2}\left|\tr U\right|^2
\ploops\label{5}\end{equation}
where $a$ and $b$ are constants of order 1. The loop effects are nonanalytic
contributions to the effective potential from quantum effects. The most
important of these is the Coleman-Weinberg term.~\cite{cw} These become very
important if the gauge couplings are very small. We can usually ignore them
because our gauge couplings are large enough that these logarithmic terms do
not change the orders of magnitudes of things, which are all we can estimate
anyway. We will not include them explicitly unless they make a difference (we
will see such an example in section \ref{tsec}).

\section{\label{ctsmsec}The CTSM Vacuum Alignment}

More formally, the low energy theory (assuming that all of the flavor fermions
have mass less than $\Lambda_2$ --- section \ref{tsec}) has an
$\su{2k+n}^4/\su{2k+n}^2$ symmetry
structure.  It is convenient to put the fermions carrying $X_U$ together into
row vectors, transforming as follows:
\begin{equation}\ba{l}
\pmatrix{[X_Uk_{LU}]_L&[X_UU]_L\cr}\rightarrow
\pmatrix{[X_Uk_{LU}]_L&[X_UU]_L\cr}\,L_{XU}^\dagger\,,\\
\pmatrix{[X_Uk_{U}]_R&[X_UL]_R\cr}\rightarrow
\pmatrix{[X_Uk_{U}]_R&[X_UL]_R\cr}\,R_{XU}^\dagger\,.\ea
\label{urows}
\end{equation}
Likewise, we can put the fermions carrying $X_D$ together into row vectors,
transforming as follows:
\begin{equation}\ba{l}
\pmatrix{[X_Dk_{LD}]_L&[X_DD]_L\cr}\rightarrow
\pmatrix{[X_Dk_{LD}]_L&[X_DD]_L\cr}\,L_{XD}^\dagger\,,\\
\pmatrix{[X_Dk_{D}]_R&[X_DL]_R\cr}\rightarrow
\pmatrix{[X_Dk_{D}]_R&[X_DL]_R\cr}\,R_{XD}^\dagger\,.\ea
\label{drows}
\end{equation}
\eqr{urows} and \eqr{drows} define the $\su{2k+n}^4$ symmetry of the strongly
interacting theory.

The vacuum condensate is described by the two $2k$+$n$$\times$$2k$+$n$ unitary
matrices, $\Sigma_U$ and $\Sigma_D$, the CCWZ~\cite{ccwz} Goldstone boson
matrices
for the symmetry breaking pattern, $\su{2k+n}^4/\su{2k+n}^2$. The condensate
$\Sigma_U$ is
\begin{equation}\ba{l}
\Sigma_U\propto
\left\langle\pmatrix{\ol{[k_UX_U]_R}\cr\ol{[LX_U]_R}\cr}
\pmatrix{[X_Uk_{LU}]_L&[X_UU]_L\cr}\right\rangle\\
=\pmatrix{\Sigma_U^{UL}&\Sigma_U^{UU}\cr\Sigma_U^{LL}&\Sigma_U^{LU}\cr}
\rightarrow R_{XU}\Sigma_UL_{XU}^\dagger\,.\ea
\label{xiu}
\end{equation}
The condensates correspond to LR fermion pairs as illustrated in
\fig{ucondensates}.
\begin{figure}$$
\bp(120,80)(-60,-30)
\put(0,0){\circle{20}}
\put(0,0){\makebox(0,0){${s\atop X_U}$}}
\put(10,0){\line(1,1){20}}
\put(30,20){\vector(-1,-1){10}}
\put(10,0){\line(1,-1){20}}
\put(10,0){\vector(1,-1){10}}
\put(-10,0){\line(-1,1){20}}
\put(-10,0){\vector(-1,1){10}}
\put(-10,0){\line(-1,-1){20}}
\put(-30,-20){\vector(1,1){10}}
\put(30,30){\circle{20}}
\put(-30,30){\makebox(0,0){${n+k\atop U}$}}
\put(30,30){\makebox(0,0){${n+k\atop L}$}}
\put(30,-30){\makebox(0,0){$k_{LU}$}}
\put(-30,-30){\makebox(0,0){$k_{U}$}}
\put(-30,30){\circle{20}}
\put(0,30){\makebox(0,0){$\Sigma_U^{LU}$}}
\put(0,-30){\makebox(0,0){$\Sigma_U^{UL}$}}
\put(40,0){\makebox(0,0){$\Sigma_U^{LL}$}}
\put(-40,0){\makebox(0,0){$\Sigma_U^{UU}$}}
\ep$$
\caption{\label{ucondensates}Condensates in $\Sigma_U$.}
\end{figure}

Likewise, for $\Sigma_D$,
\begin{equation}\ba{l}
\Sigma_D\propto
\left\langle\pmatrix{\ol{[k_DX_D]_R}\cr\ol{[LX_D]_R}\cr}
\pmatrix{[X_Dk_{LD}]_L&[X_DD]_L\cr}\right\rangle\\=
\pmatrix{\Sigma_D^{DL}&\Sigma_D^{DD}\cr\Sigma_D^{LL}&\Sigma_D^{LD}\cr}
\rightarrow R_{XD}\Sigma_DL_{XD}^\dagger\ea
\label{xid}
\end{equation}
These correspond to LR fermion pairs as illustrated in \fig{dcondensates}.
\begin{figure}$$
\bp(120,80)(-60,-30)
\put(0,0){\circle{20}}
\put(0,0){\makebox(0,0){${s\atop X_D}$}}
\put(10,0){\line(1,1){20}}
\put(10,0){\vector(1,1){10}}
\put(10,0){\line(1,-1){20}}
\put(30,-20){\vector(-1,1){10}}
\put(-10,0){\line(-1,1){20}}
\put(-30,20){\vector(1,-1){10}}
\put(-10,0){\line(-1,-1){20}}
\put(-10,0){\vector(-1,-1){10}}
\put(30,30){\circle{20}}
\put(-30,30){\makebox(0,0){${n+k\atop L}$}}
\put(30,30){\makebox(0,0){${n+k\atop D}$}}
\put(30,-30){\makebox(0,0){$k_{D}$}}
\put(-30,-30){\makebox(0,0){$k_{LD}$}}
\put(-30,30){\circle{20}}
\put(0,30){\makebox(0,0){$\Sigma_D^{LD}$}}
\put(0,-30){\makebox(0,0){$\Sigma_D^{DL}$}}
\put(40,0){\makebox(0,0){$\Sigma_D^{DD}$}}
\put(-40,0){\makebox(0,0){$\Sigma_D^{LL}$}}
\ep$$
\caption{\label{dcondensates}Condensates in $\Sigma_D$.}
\end{figure}

In the CTSM vacuum alignment, we have
\begin{equation}
\Sigma_U^{UL}=0\,,\quad
\Sigma_U^{UU}=\pmatrix{I&0\cr}\,,\quad
\Sigma_U^{LL}=\pmatrix{-I\cr0\cr}\,,\quad
\Sigma_U^{LU}=\pmatrix{0&0\cr0&I\cr}\,,
\label{uvevs}
\end{equation}
\begin{equation}
\Sigma_D^{DL}=0\,,\quad
\Sigma_D^{DD}=\pmatrix{I&0\cr}\,,\quad
\Sigma_D^{LL}=\pmatrix{-U\cr0\cr}\,,\quad
\Sigma_D^{LD}=\pmatrix{0&0\cr0&I\cr}\,.
\label{dvevs}
\end{equation}
These condensates define the technicolor directions and the flavor directions
in the various spaces. In particular, we have broken up the $n+k$ dimensional
spaces of the ETC couplings into a $k$ dimensional flavor space and an $n$
dimensional technicolor space. Putting this together gives
\begin{equation}
\Sigma_U=\pmatrix{0&I&0\cr -I&0&0\cr 0&0&I\cr}\,,\quad
\Sigma_D=\pmatrix{0&I&0\cr -U&0&0\cr 0&0&I\cr}\,.
\label{threebythreevevs}
\end{equation}

In this basis the flavor symmetry breaking mass terms have the
form and transformation properties
\begin{equation}
\ba{c}
\BM_U=\pmatrix{\cM_U&0\cr0&0\cr}
\rightarrow L_{XU}\BM_UR_{XU}\,,\\
\BM_D=\pmatrix{\cM_D&0\cr0&0\cr}
\rightarrow L_{XD}\BM_DR_{XD}\,,\ea
\label{masses}
\end{equation}
in the relevant spaces. The generators of the various $\su{n+k}$ gauge
symmetries transform as
\begin{equation}
\ba{cl}
\cT^U_a\rightarrow L_{XU}\cT^U_a L_{XU}^\dagger&\mbox{for $U$}\\
\cT^D_a\rightarrow L_{XD}\cT^D_a L_{XD}^\dagger&\mbox{for $D$}\\
\cT^L_a\rightarrow R_{XU}\cT^L_a R_{XU}^\dagger&\mbox{and}\\
\cT^L_a\rightarrow R_{XD}\cT^L_a R_{XD}^\dagger&\mbox{for $L$}
\ea
\label{gauges}
\end{equation}
The $L$ generators transform in two different ways because they couple to both
$X_U$ and $X_D$ fermions.
All of these gauge generators have the form
\begin{equation}
\cT_a=\pmatrix{0&0\cr0&T_a\cr}
\label{gaugeform}\end{equation}
where $T_a$ are conventional $\su{n+k}$ generators.

Let us now examine the various terms in the potential. First consider the
terms that are responsible for the CTSM alignment.
The basic assumption of CTSM is that the extended technicolor gauge
interactions prefer not to combine into a diagonal sum. The relevant
invariants have the form (see \eqr{xiu},\eqr{xid} and \eqr{gauges}):
\begin{equation}
16\pi^2f_2^4\,c\,\left[{\xi_U^2\xi_L^2\over 16\pi^2}\tr
\left({\cT^U_a}^2\Sigma_U^\dagger{\cT_b^L}^2\Sigma_U\right)
+{\xi_D^2\xi_L^2\over 16\pi^2}\tr
\left({\cT^D_a}^2\Sigma_D^\dagger{\cT_b^L}^2\Sigma_D\right)\right]
\ploops\label{ctsm}
\end{equation}
where the coefficient, $c$ is order one and {\bf positive.} If this is the
case, then the vacuum aligns to minimize the positive definite form in
\eqr{ctsm}. This gives \eqr{uvevs} and \eqr{dvevs}. Of course, we do not know
what $c$ is~\cite{luty}, so one might worry that the whole idea of CTSM is
flawed if strongly interacting theories always produce the wrong alignment for
some reason. However in this model, and probably in any realistic model, there
are additional contributions from 4-fermion operators at the scale, $f_1$,
which tend to produce the desired alignment. This is easy to see in
\fig{highmoose}. If the $X_U$ and $X_D$ groups got strong before $L_{TC}$,
they
would automatically produce the CTSM alignment. Thus the 4-fermion operators
produced by these interactions when they are broken also go in the right
direction. If necessary, the gauge couplings at the $f_1$ scale could be tuned
slightly close to their critical values to make these effects large. The CTSM
alignment is not the difficult problem with CTSM.

However, even if $c>0$, the term \eqr{ctsm} must compete with the leading
effects of the mass terms, which look like (from \eqr{xiu}, \eqr{xid} and
\eqr{masses})
\begin{equation}
4\pi f_2^3 \,d \,\left[\tr(\BM_U\Sigma_U)+\tr(\BM_D\Sigma_D) +\hc\right]
\label{massterm}
\end{equation}
Unless \eqr{3} is satisfied, this term forces the left- and right-handed
components of the heavy flavor fermions to condense, destroying \eqr{uvevs}.
The situation in which \eqr{3} is satisfied for everything except the $t$
quark is especially interesting, and we will discuss it explicitly below.

\begin{figure}$$
\bp(300,150)(-75,-65)
\multiput(0,0)(150,0){2}{\ptcommand{\circulararc 360 degrees from 50 0 center
at 0 0}\put(0,50){\makebox(0,0){\large$\times$}}\put(0,-50)
{\makebox(0,0){\large$\times$}}}
\put(0,60){\makebox(0,0){$\Sigma_U^\dagger$}}
\put(0,-60){\makebox(0,0){$\Sigma_U$}}
\put(150,60){\makebox(0,0){$\Sigma_D$}}
\put(150,-60){\makebox(0,0){$\Sigma_D^\dagger$}}
\put(40,30){\photonru{7}}
\put(40,-30){\photonru{7}}
\put(75,20){\makebox(0,0){$L$}}
\put(75,-20){\makebox(0,0){$L$}}
\put(-40,30){\makebox(0,0){\large +}}
\put(-40,-30){\makebox(0,0){\large +}}
\put(190,30){\makebox(0,0){\large +}}
\put(190,-30){\makebox(0,0){\large +}}
\put(-55,30){\makebox(0,0){$\BM_U$}}
\put(-55,-30){\makebox(0,0){$\BM_U$}}
\put(-45,0){\makebox(0,0)[l]{$[X_Uk_U]_R$}}
\put(-20,50){\makebox(0,0)[rb]{$[X_Uk_{LU}]_L$}}
\put(-20,-50){\makebox(0,0)[rt]{$[X_Uk_{LU}]_L$}}
\put(20,50){\makebox(0,0)[lb]{$[X_UL]_R$}}
\put(20,-50){\makebox(0,0)[lt]{$[X_UL]_R$}}
\put(205,30){\makebox(0,0){$\BM_D$}}
\put(205,-30){\makebox(0,0){$\BM_D$}}
\put(195,0){\makebox(0,0)[r]{$[X_Dk_D]_R$}}
\put(170,50){\makebox(0,0)[lb]{$[X_Dk_{LD}]_L$}}
\put(170,-50){\makebox(0,0)[lt]{$[X_Dk_{LD}]_L$}}
\put(130,50){\makebox(0,0)[rb]{$[X_DL]_R$}}
\put(130,-50){\makebox(0,0)[rt]{$[X_DL]_R$}}
\ep
$$\caption{\label{moose12}Feynman graph contributing to \protect\eqr{u1}.}
\end{figure}

\section{The KM Matrix\label{kmsec}}

For now we will assume that \eqr{3} is satisfied and that the condensates have
the general CTSM form, \eqr{uvevs} and \eqr{dvevs}, and consider the vacuum
alignment problem for $U$. The relevant terms must involve both $X_U$ and
$X_D$ fermions (otherwise the $U$ dependence could be rotated away). One such
contribution involves the exchange of two $L$ gauge bosons between the two
types of fermions, as shown in \fig{moose12}. This can lead to a term
proportional to
\begin{equation}
\tr\left(\Sigma_U\BM_U\BM_U^\dagger\Sigma_U^\dagger\cT_a^L\cT_b^L\right)
\tr\left(\Sigma_D\BM_D\BM_D^\dagger\Sigma_D^\dagger\cT_b^L\cT_a^L\right)
\label{u1}
\end{equation}
which contains the first term in \eqr{5}.

\begin{figure}$$
\bp(130,150)(-65,-65)
\put(0,0){\ptcommand{\circulararc 360 degrees from 50 0 center at 0 0}}
\put(2,50){\photondr{10}}
\put(-2,50){\photondr{10}}
\put(-50,0){\photonrd{10}}
\put(-40,30){\makebox(0,0){\large +}}
\put(-40,-30){\makebox(0,0){\large +}}
\put(40,30){\makebox(0,0){\large +}}
\put(40,-30){\makebox(0,0){\large +}}
\put(-55,0){\makebox(0,0)[r]{$[X_UL]_R$}}
\put(55,0){\makebox(0,0)[l]{$[X_DL]_R$}}
\put(-50,30){\makebox(0,0)[r]{$\Sigma_U$}}
\put(-50,-30){\makebox(0,0)[r]{$\Sigma_U^\dagger$}}
\put(50,30){\makebox(0,0)[l]{$\Sigma_D^\dagger$}}
\put(50,-30){\makebox(0,0)[l]{$\Sigma_D$}}
\put(-20,50){\makebox(0,0)[rb]{$[X_Uk_{LU}]_L$}}
\put(-20,-50){\makebox(0,0)[rt]{$[X_Uk_{LU}]_L$}}
\put(20,50){\makebox(0,0)[lb]{$[X_Dk_{LD}]_L$}}
\put(20,-50){\makebox(0,0)[lt]{$[X_Dk_{LD}]_L$}}
\ep
$$\caption{\label{moose13}Feynman graph contributing to \protect\eqr{u2}.}
\end{figure}

Another way to get nontrivial $U$ dependence is to make use of the 4-fermion
operators of \eqr{2} from the $f_1$ scale. Again, the $L$ gauge couplings must
get into the act as well (although here, one gauge boson exchange is
sufficient). The leading term comes from an operator proportional to (see
\fig{moose13})
\begin{equation}
\tr\left(\Sigma_UP_k\Sigma_D^\dagger\cT_a^L\Sigma_DP_k\Sigma_U^\dagger\cT_a^L
\right)\,,
\label{u2}
\end{equation}
where $P_k$ is a projection operator,
\begin{equation}
P_k=\pmatrix{I&0\cr0&0\cr}\,.
\label{pk}
\end{equation}
\eqr{u2} contains the second term in \eqr{5}.

To see what \eqr{5} implies, choose the basis in which
\begin{equation}\cM_U=M_U\,,\quad\cM_D=S\,M_D\,,\label{10}\end{equation}
where
$M_U$ and $M_D$ are diagonal and $S$ is unitary. As we will see explicitly
below, the unitary matrix, $V=US$ is
the KM matrix, at least in lowest order. The matrix $S$ is fixed by the high
energy physics. The KM matrix, $V$ is then fixed by the vacuum alignment.

In this basis, the potential \eqr{5} has the form
\begin{equation} -{a\over 16\pi^2}{\xi_L^2}\tr\left(M_U^2 V
M_D^2 V^\dagger\right)
-{4\pi\xi_Lf_2^6b\over f_1^2}\left|\tr \left(VS^\dagger\right)
\right|^2\ploops\,.\label{12}\end{equation}
If $a$ and $b$ are positive, then the first term in \eqr{12} tends to drive
$V\rightarrow
I$ in the basis in which the masses are in order, while the second term tends
to drive $V\rightarrow S$. The question is what happens when the largest
masses are large enough to cause the first term to dominate, while the light
masses are small enough for the second to dominate.

We will minimize \eqr{12}, even though loops will also contribute, just to
show how the calculation goes.
It is straightforward to minimize \eqr{12} (or its generalization to include
loop effects) in the limit in which we neglect
all of the mass terms except those proportional to $m_t^2m_b^2$.  This is
probably a reasonable approximation because the next largest term is smaller
by a factor of $m_s^2/m_b^2$, which is pretty huge. Then we can rewrite
\eqr{12} as
\begin{equation} -A\left|V_{tb}\right|^2-B\left|\tr(VS^\dagger)\right|^2\,.
\label{13}\end{equation}
An arbitrary 3$\times$3 matrix, $V$ can be parameterized as follows (with
$V_{tb}=V_{33}$):
\begin{equation}
V=\pmatrix{e^{i(\beta-\alpha)}\left(1-(1-c_\phi)uu^\dagger\right)\Sigma &-
s_\phi u\cr
e^{i\beta}s_\phi u^\dagger\Sigma&e^{i\alpha}c_\phi\cr}\,,
\label{14}\end{equation}
where $s_\phi$ ($c_\phi$) is $\sin\phi$ ($\cos\phi$), $\alpha$ and $\beta$ are
arbitrary phases, $\Sigma$ is a unitary, unimodular 2$\times$2 matrix, and $u$
is an arbitrary two component complex vector.\myfoot{Note that this trick can
be applied to an $N$$\times$$N$ unitary matrix, to write it in terms of an
$N$$-$1 vector, and $N$$-$1$\times$$N$$-$1 unitary matrix, and phases.} The
point of this is
that we can
parameterize $S$ in the same way:
\begin{equation}
S=\pmatrix{e^{i(\beta'-\alpha')}\left(1-(1-c_{\phi'}){u'}
{u'}^\dagger\right)\Sigma'&- s_{\phi'} {u'}\cr
e^{i\beta'}s_{\phi'} {u'}^\dagger\Sigma'&e^{i\alpha'}c_{\phi'}\cr}\,.
\label{14'}\end{equation}
Then the second term in \eqr{13} is clearly extremized for
\begin{equation} \alpha=\alpha'\,,\quad \beta=\beta'\,,\quad
\Sigma=\Sigma'\,,\quad u=u'\,. \label{15}\end{equation}
Then
\begin{equation}
U=VS^\dagger=\pmatrix{1-uu^\dagger+\cos(\phi-\phi')uu^\dagger
&-e^{-i\alpha}\sin(\phi-\phi')u\cr
e^{i\alpha}\sin(\phi-\phi')u^\dagger&\cos(\phi-\phi')\cr}
\label{vsdagger}
\end{equation}
and \eqr{13} becomes
\begin{equation} -A\cos^2\phi-B\left[1+2\cos(\phi-\phi')\right]^2 \,.
\label{phipot}\end{equation}
While we don't know anything about $\phi'$, we know experimentally that $\phi$
is very small. Thus we can get an excellent approximation to the minimum by
expanding \eqr{phipot} to second order in $\phi$. The result for the minimum
is
\begin{equation}
\phi\approx{2B\sin\phi'(1+2\cos\phi')\over
A+2B\cos\phi'+4B\cos2\phi'}\,.
\label{phiapprox}
\end{equation}
Note that the loop effects that we have ignored will change the details of
\eqr{phiapprox}, but not the rough sizes of the terms. We will use this below.

\section{ETC Interactions\label{etcsec}}

The light fermions communicate with the flavor physics through
the various extended technicolor interactions. Our next job is to write down
the masses of the ETC gauge bosons.
The largest terms come from the kinetic energy terms in the chiral Lagrangians
for $\Sigma_U$ and $\Sigma_D$,
\begin{equation}
{f_2^2\over4}\left[\tr\left(D^\mu\Sigma_UD_\mu\Sigma_U^\dagger\right)+
\tr\left(D^\mu\Sigma_DD_\mu\Sigma_D^\dagger\right)\right]\,,
\label{keterm}
\end{equation}
where the covariant derivative is
\begin{equation}
\ba{c}
D^\mu\Sigma_U=\partial^\mu\Sigma_U+ig_L\cT_a^LW_{La}^\mu\Sigma_U
-ig_L\Sigma_U\cT_a^UW_{Ua}^\mu\,,\\
D^\mu\Sigma_D=\partial^\mu\Sigma_D+ig_L\cT_a^LW_{La}^\mu\Sigma_D
-ig_L\Sigma_D\cT_a^DW_{Da}^\mu\,.\ea
\label{covariant}
\end{equation}
The ETC masses arising from \eqr{keterm} do not give light fermion masses. The
leading mass terms arise from the terms
\begin{equation}
{af_2\over16\pi}\left[
\tr\left(\Sigma_U\BM_UD^\mu\Sigma_UD_\mu\Sigma_U^\dagger\right)+
\tr\left(\Sigma_D\BM_DD^\mu\Sigma_DD_\mu\Sigma_D^\dagger\right)+\hc\right]\,,
\label{kemass}
\end{equation}
This gives the standard CTSM masses.

\begin{figure}
$$
\bp(80,80)(-40,-30)
\put(-40,20){\line(1,0){80}}
\put(-40,20){\vector(1,0){20}}
\put(0,20){\vector(1,0){20}}
\put(-40,-20){\line(1,0){80}}
\put(40,-20){\vector(-1,0){20}}
\put(0,-20){\vector(-1,0){20}}
\put(-2,20){\photondr{4}}
\put(2,20){\photondr{4}}
\put(-20,25){\makebox(0,0)[rb]{$[LX_U]_L$}}
\put(-20,-25){\makebox(0,0)[rt]{$[LX_U]_L$}}
\put(20,25){\makebox(0,0)[lb]{$[LX_D]_L$}}
\put(20,-25){\makebox(0,0)[lt]{$[LX_D]_L$}}
\ep
$$
\caption{\label{moose15}Feynman graph contributing to \protect\eqr{fcnc1}.}
\end{figure}

Now consider FCNC effects from the term that arises (for example) from
\fig{moose15}.
The exchange of the heavy vector bosons at the scale $f_1$ in \fig{moose15}
gives rise to 4-fermion interactions that we can write in the following form
(after Fierz transformation)
\begin{equation}
{1\over f_1^2}\tr\left(\ol{[X_UL]_L}T_a\gamma^\mu[LX_U]_L\right)
\tr\left(\ol{[X_DL]_L}T_a\gamma_\mu[LX_D]_L\right)\,,
\label{fcnc0}
\end{equation}
where $T_a$ are the conventionally normalized Gell-Mann matrices for
$SU(n+k)\times U(1)$. The normalization of \eqr{fcnc0} amounts to a definition
of the scale $f_1$. However, the sign of \eqr{fcnc0} is meaningful. It is
determined by the fact that the terms comes from massive vector boson
exchange.

\eqr{fcnc0} has the form of a sum of products of a current in $X_U$ sector of
the theory times a current in $X_D$. Below the symmetry breaking scale,
$\Lambda_2$, the form of these currents is fixed in leading order in the
derivative expansion by Noether's theorem (for the $SU(n+k)$
currents\myfoot{The $U(1)$ currents go to zero in leading order.}):
\begin{equation}
\begin{array}{l}
\tr\left(\ol{[X_UL]_L}T_a\gamma^\mu [LX_U]_L\right)
\rightarrow i{f_2^2\over 2}\tr\left(
\Sigma_U^\dagger T_a D^\mu \Sigma_U\right)\\
\tr\left(\ol{[X_DL]_L}T_a\gamma^\mu [LX_D]_L\right)
\rightarrow i{f_2^2\over 2}\tr\left(
\Sigma_D^\dagger T_a D^\mu \Sigma_D\right)\,.\end{array}
\label{fcnc01}
\end{equation}

This gives rise to a mass term for the $L$ gauge bosons of the
form
\begin{equation}
{g_L^2f_2^4\over 8f_1^2}\tr\left(\cT_a^LW_{La}^\mu \Sigma_U\Sigma_D^\dagger
\cT_b^LW_{Lb\mu} \Sigma_D\Sigma_U^\dagger\right)
\label{fcnc1}
\end{equation}
For the flavor generators,
\begin{equation}
\cT_a^L=\pmatrix{0&0&0\cr0&t_j&0\cr0&0&0\cr}\,,
\label{flavor}
\end{equation}
This gives a term in the mass matrix
\begin{equation}
\Delta m_{jk}^2=
{g_L^2f_2^4\over 16f_1^2}\tr\left(t_jUt_kU^\dagger\right)+\hc
\label{flavormass}
\end{equation}
which in turn gives rise to FCNC of the following form (through the diagram of
figure \ref{moose152}, again with appropriate Fierz transformations):
\begin{equation}
{1\over f_1^2}\,\left(\ol{\psi_L}\gamma^\mu U\psi_L\right)
\,\left(\ol{\psi_L}\gamma_\mu U^\dagger\psi_L\right)
\label{leftfcnc}
\end{equation}
where $\psi_L$ is the left-handed quark doublet,
\begin{equation}
\psi_L=\pmatrix{u\cr Vd\cr}\,.
\label{quarks}
\end{equation}
For the charge 2/3 quarks, this looks like
\begin{equation}
{1\over f_1^2}\,\left(\ol{u_L}\gamma^\mu VS^\dagger u_L\right)
\,\left(\ol{u_L}\gamma_\mu SV^\dagger u_L\right)\,.
\label{ufcnc}
\end{equation}
For the charge $-$1/3 quarks, it looks like
\begin{equation}
{1\over f_1^2}\,\left(\ol{d_L}\gamma^\mu S^\dagger Vd_L\right)
\,\left(\ol{d_L}\gamma_\mu V^\dagger Sd_L\right)\,.
\label{dfcnc}
\end{equation}
Note that this depends on the combination
\begin{equation}
S^\dagger V=
\pmatrix{1-vv^\dagger+\cos(\phi-\phi')vv^\dagger
&e^{-i\alpha}\sin(\phi-\phi')v\cr
-e^{i\alpha}\sin(\phi-\phi')v^\dagger&\cos(\phi-\phi')\cr}
\label{sdaggerv}
\end{equation}
where
\begin{equation}
v=\Sigma^\dagger u\,.
\label{vsigmau}
\end{equation}
The potentially interesting thing about all this is that it is a FCNC effect
that has a form slightly different from the usual box graph and its CTSM
generalizations. This may make it possible to find evidence for the existence
of the $f_1$ scale by studying low energy FCNC processes.\myfoot{Work in the
direction is in progress with C. Carone and R. Hamilton.}
\begin{figure}
$$
\bp(120,70)(-60,-30)
\thicklines
\put(-40,-20){\line(0,1){40}}
\put(40,-20){\line(0,1){40}}
\put(-40,0){\photonru{8}}
\put(0,0){\makebox(0,0){\bf +}}
\put(0,12){\makebox(0,0){$\Delta m_{jk}^2$}}
\put(-50,0){\makebox(0,0){$g_L$}}
\put(50,0){\makebox(0,0){$g_L$}}
\put(-20,-17){\makebox(0,0){$\displaystyle 1\over\displaystyle g_L^2f_2^2$}}
\put(20,-17){\makebox(0,0){$\displaystyle 1\over\displaystyle g_L^2f_2^2$}}
\put(-40,30){\makebox(0,0){$\psi_L$}}
\put(40,30){\makebox(0,0){$\psi_L$}}
\ep
$$
\caption{\label{moose152}Feynman graph contributing to
\protect\eqr{leftfcnc}.}
\end{figure}

\section{Heavy $t$\label{tsec}}

Now consider what happens if the $t$ type flavor fermions is too heavy. I
don't think that it makes much difference whether it is too heavy for \eqr{3}
to be satisfied or if it is so heavy that it doesn't exist in the theory at
$f_2$ at all. I will assume the latter. Then many of the formulas in the
previous section remain the same but their meaning changes. In particular, in
\eqr{threebythreevevs}, the first two spaces are now $k-3$ dimensional, while
the last is $n+3$ dimensional.

\begin{figure}$$
\bp(185,150)(25,-65)
\put(150,0){\ptcommand{\circulararc 360 degrees from 50 0 center
at 0 0}\put(0,50){\makebox(0,0){\large$\times$}}\put(0,-50)
{\makebox(0,0){\large$\times$}}}
\put(150,60){\makebox(0,0){$\Sigma_D$}}
\put(150,-60){\makebox(0,0){$\Sigma_D^\dagger$}}
\put(40,30){\photonru{7}}
\put(40,-30){\photonrd{7}}
\put(40,30){\photondl{6}}
\put(40,30){\makebox(0,0){\large +}}
\put(40,-30){\makebox(0,0){\large +}}
\put(30,0){\makebox(0,0){$U$}}
\put(75,20){\makebox(0,0){$L$}}
\put(75,-20){\makebox(0,0){$L$}}
\put(190,30){\makebox(0,0){\large +}}
\put(190,-30){\makebox(0,0){\large +}}
\put(205,30){\makebox(0,0){$\BM_D$}}
\put(205,-30){\makebox(0,0){$\BM_D$}}
\put(195,0){\makebox(0,0)[r]{$[X_Dk_D]_R$}}
\put(170,50){\makebox(0,0)[lb]{$[X_Dk_{LD}]_L$}}
\put(170,-50){\makebox(0,0)[lt]{$[X_Dk_{LD}]_L$}}
\put(130,50){\makebox(0,0)[rb]{$[X_DL]_R$}}
\put(130,-50){\makebox(0,0)[rt]{$[X_DL]_R$}}
\ep
$$\caption{\label{moose16}Feynman graph contributing to \protect\eqr{u1t}.}
\end{figure}

The main difference in this situation is that we don't need factors of $\cM_t$
to distinguish the $t$ direction in flavor space. The structure of the
condensate already picks out the $t$ direction. The estimates of the previous
section remain in force with $\cM_t$ replaced by $\Lambda_2$. For example, the
analog of \fig{moose12} is the diagram in \fig{moose16}. This contributes to a
term proportional to
\begin{equation}
\tr\left(\Sigma_U\cT^U_c\Sigma_U^\dagger\cT_a^L\right)
\tr\left(\Sigma_U\cT^U_c\Sigma_U^\dagger\cT_b^L\right)
\tr\left(\Sigma_D\BM_D\BM_D^\dagger\Sigma_D^\dagger\cT_b^L\cT_a^L\right)
\label{u1t}
\end{equation}
This looks like it is of order $\xi_U\xi_L^2$. However, because the loop
integral over the vector boson momentum is rapidly convergent, it is actually
of order $\Sigma^2\times\mbox{logs}$.\myfoot{The leading term is the
Coleman-Weinberg contribution to the effective potential. This is one of the
loop effects that we have been ignoring. Here it makes a difference.} It
contributes to
the first term in the
potential that determines $U$, which now looks roughly like
\begin{equation} -{af_2^2}{\Sigma^2}
\tr\left(P_t U\cM_D\cM_D^\dagger U^\dagger\right)
-{\xi_L}{16\pi^2f_2^6b\over f_1^2}\left|\tr U\right|^2
\label{5t}\end{equation}
where $P_t$ is a projection operator onto the $t$ direction.
As promised, this is essentially equivalent to \eqr{5}, with
$\cM_t\approx\Lambda_2$.

\section{Scales\label{scalesec}}

Assuming
\begin{equation}
\cM_t\approx\Lambda_2\,,
\label{ta}
\end{equation}
we can use improved NDA to estimate the various scales in the problem.
Begin with the CTSM relation
\begin{equation}
m_q\approx{\Lambda v^2\over \Lambda_2f_2^2}\cM\,.
\label{ctsmrelation}
\end{equation}
where, $v$ and $\Lambda$ are the decay constant and mass for the technicolor
scale and $f_2$ and $\Lambda_2$ are the corresponding constants for the
ultracolor scale.
Then \eqr{ta} implies
\begin{equation}
f_2^2\approx v^2{\rho v\over m_t}
\label{ctsm2}
\end{equation}
where $\rho$ is the ratio
\begin{equation}
\rho\equiv{\Lambda\over v}\,.
\label{rho}
\end{equation}
We will also define the corresponding ratios for the other scales in the
problem,
\begin{equation}
\rho_i\equiv{\Lambda_i\over v_i}\;\;\forto i02\,.
\label{rhoi}
\end{equation}
We expect from the arguments of \cite{weinberg,nda} that
\begin{equation}
\rho\,,\;\rho_i\lta4\pi\,.
\label{rhobound}
\end{equation}

Note also that \eqr{ctsm2} puts a strong bound on $m_t$. With
$v\approx250$~Gev, $\rho\approx4\pi$, and $m_t\approx90$~GeV,~\cite{tbound}
\eqr{ctsm2} gives $f_2\approx1.5$~TeV. We cannot tolerate $f_2$ much smaller
than that.~\cite{comp} If the $t$ is much heavier than that, some walking or
other enhancement mechanism would be required to keep $f_2$ large enough.

We can now fix the scale, $f_1$.
\begin{equation}
A\approx f_2^2 \xi_L^2 \cM_b^2\approx {\Lambda_2^2f_2^6m_b^2\xi_L^2\over
\Lambda^2v^4}\,,
\label{a}
\end{equation}
In \eqr{a} we have assumed \eqr{5t} (or \eqr{5} with $\cM_t\approx\Lambda_2$).
we have also used the CTSM relation Likewise
\begin{equation}
B\approx {\Lambda_2^2f_2^4\xi_L\over f_1^2}\,.
\label{b}
\end{equation}
Then \eqr{phiapprox} tells us (note that I am only using the approximate
statement, so the neglect of loop effects doesn't hurt)
\begin{equation}
A\phi\approx B\phi'\,.
\label{phiapprox2}
\end{equation}
Now, it is interesting to see how large $f_1$ can be. We see from
\eqr{a}-\eqr{phiapprox2} that the way to do this is to make the angle $\phi'$
large. If we take it of order one, we have $A\phi\approx B$, or
\begin{equation}
f_1^2\approx{1\over\phi\xi_L}\,\Lambda^2\,{v^2\over f_2^2}\,
{v^2\over m_b^2}\approx v^2{\rho\over \phi\xi_L}
{m_t\over v}{v^2\over m_b^2}\,.
\label{fone}
\end{equation}

Finally, we can obtain a bound on the scale $f_0$ by considering the
ultrafermion masses, \eqr{fermionmasses}, that are produced by the 4-fermion
operators, \eqr{fourfermionmasses}. The largest that one of these can be is
\begin{equation}
{\Lambda_1f_1^2\over f_0^2}\,.
\label{f0mass}
\end{equation}
To give the largest possible $t$ mass, we want one of these to satisfy
(see \eqr{3})
\begin{equation}
\cM_t\gta\xi_L\xi_U\Lambda_2\,.
\label{f0tmass}
\end{equation}
Putting these together gives a bound on $f_0$,
\begin{equation}
f_0^2\lta f_1^2{\rho_1\over \xi_L\xi_U\rho_2}{f_1\over f_2}\,.
\label{f0bound}
\end{equation}

\section{Conclusions\label{lastsec}}
While this model is ugly, it has some very beautiful features, for example:
\begin{itemize}
\item the unusual mechanism for the generation of a large $t$ mass;
\item the interplay
between the masses and the explicit symmetry breaking that determines the KM
angle;
\item the possibility of the natural appearance of a custodial $SU(2)$
symmetry.
\end{itemize}
But the most interesting feature of all is that the scales involved are quite
reasonable. For example, if we assume $m_t\approx100$~GeV and $\xi_U$,
$\xi_D$, $\xi_L$, and $\phi$ are all approximately 1, and  $\rho$, $\rho_1$
and $\rho_2\approx4\pi$, we get
\begin{equation}
f_2\approx 1.5\;\mbox{TeV}\,,\quad
f_1\approx 140\;\mbox{TeV}\,,\quad
f_0\approx 1300\;\mbox{TeV}\,.
\label{scaleestimate}
\end{equation}
The scale $f_0$, above which there are unspecified GIM violating interactions,
is actually comfortably large without fine tuning or walking.\myfoot{Compare
this with recent attempts to use walking, for example the model of Sundrum
\cite{sundrum}, in which there are GIM violating interactions at lower
scales.} Note however that the estimate of $f_1$ is well below the upper bound
discussed in \cite{ctsmquarks} (from other considerations). The physics that
determines the KM matrix leaves the
characteristic FCNC, (\ref{ufcnc}) and (\ref{dfcnc}),
at the scale $f_1$ below $f_0$. These interactions are distinctive signals of
CTSM (or ETSM) models. If they can be seen, they may be the first hint of the
ultimate breakdown of the GIM mechanism that must occur at very high energy.

\section*{Acknowledgements}
I am grateful to Lisa Randall for many insightful comments about these
models, and to Chris Carone for careful reading of the manuscript and for
useful discussions of the phenomenology. Research supported in part by the
National Science Foundation under Grant \#PHY-8714654 and by the Texas
National Research Laboratory Commission, under Grant \#RGFY9206.

\end{document}